\documentclass[preprint,aps,psfig,showpacs]{revtex4}

\usepackage{graphicx}

\begin{document}
\title{Nonclassical degree of states of single and bipartite systems}
\author{A. T. Avelar$^{\;a,b}$, B. Baseia$^{\;a}$
and J. M. C. Malbouisson$^{\;c}$}
\address{$^{a}$Instituto de F\'{i}sica, Universidade Federal
de Goi\'{a}s, 74.001-970, Goi\^{a}nia, GO, Brazil\\
$^{b}$Instituto de F\'{i}sica, Universidade de
Bra\'{i}lia, 70.919-970, Bra\'{i}lia, DF, Brazil\\
$^{c}$Instituto de F\'{i}sica, Universidade Federal da Bahia,
40.210-340, Salvador, BA, Brazil}

\begin{abstract}
We consider experimental routes to determine the nonclassical
degree of states of a field mode. We adopt a distance-type
criterium based on the Hilbert-Schmidt metric to quantify the
nonclassicality. The concept of nonclassical degree is extended
for states of bipartite systems, allowing us to discuss a possible
connection between nonclassicality and entanglement measures.
\end{abstract}


\maketitle

There are several interesting states of the quantized
electromagnetic field studied nowadays, either concerning with
their properties or their creation in laboratories. They are
considered as classical states when their Glauber-Sudarshan
P-functions are regular, non-negative; otherwise, when one of such
characteristics are not attained, they are said to be nonclassical
\cite{Glauber63}. Coherent states and mixed thermal states are
representative examples of classical states. According to a
theorem by Hillery \cite{Hillery85}, every pure field state which
is not coherent, is nonclassical. This result leads us to an
endless number of nonclassical states in quantum optics. These
states exhibit quantum effects, the most traditional of them
being: (i) antibunching \cite{Kimble77}, (ii) sub-Poissonian
statistics \cite{Mandel83} and (iii) squeezing \cite{Stoler70}.
There are even other examples of field states which do not exhibit
these quantum effects, but their field quantization are required
to explain some experimental results. One such situation occurs
for the time evolution of atomic inversion when the atom interacts
with a field inside a cavity: the collapse-revival effect
\cite{ENM} can be explained only when the field is quantized, even
if it is in the (most classical) pure coherent state.

Traditional examples of nonclassical states of the radiation field
are: (i) the well known number (Fock) state $\left| n\right\rangle
$, exhibiting antibunching and maximum sub-Poissonian; (ii) the
squeezed coherent state $ \left| z,\alpha \right\rangle ,$\ which
may be sub-Poissonian or not, depending on the type of squeezing
effect; (iii) the phase-state \cite {Pegg88}, which is
nonclassical according to the Hillery's theorem \cite {Hillery85},
but it does not exhibit any known quantum effect, a result which
stimulated the investigation about the nonclassical depth of this
state \cite{Marchiolli01}.

However, quantum properties do not occur simultaneously for all
nonclassical states. For example, all squeezed-vacuum states are
super-Poissonian while squeezed-coherent states may be
sub-Poissonian. Also, all number states show maximum
sub-Poissonian statistics but exhibit no squeezing and their
antibunching effect diminishes when N increases \cite{Walls84}.

The above considerations lead to the appropriated question about
how much nonclassical a quantum state is. Various criteria have
appeared in the literature to quantify the nonclassical character
of a given state. One such trial was introduced by Mandel
\cite{Mandel79}, defining the parameter $q=\left( \Delta
\widehat{n}^{2}-\left\langle \widehat{n}\right\rangle \right)
/\left\langle \widehat{n}\right\rangle $ to quantify the departure
of the photon-number distribution of the state from the Poissonian
statistics. However, this parameter is not able to contemplate
other nonclassical properties. A proposal by Hillery \cite
{Hillery89} introduced a nonclassical distance of a state as the
trace-norm of the difference between the density operator of the
state and that of the nearest classical state. However, practical
determination of such distance is rather difficult. Following this
trend, other more operational measures of nonclassicality were
introduced: one by Dodonov et al \cite{Dodonov99}, via a
Hilbert-Schmidt distance between density operators; another by
Marian et al \cite{Marian02}, via the Bures-Uhlmann definition of
distance between states. Measure of nonclassical properties
\cite{Kim} and observable criterium distinguishing nonclassical
states were also considered recently \cite{Vogel00}.

A distinct route, based on the Cahill-Glauber representation
\cite{Glauber69}, was early proposed by Lee \cite{Lee91},
introducing the R-function as a real $ \tau $-parametrized
Gaussian convolution of the P-function. With this representation,
Lee defined the nonclassical depth of a state as the minimum value
of $\tau $ ($\tau _{m}$) yielding to a regular, non-negative,
R-function acceptable as a classical distribution function. Later
on, Lutkenhaus and Barnett \cite{Barnett95} considered a similar
phase-space measure of nonclassicality. Very recently
\cite{Malbouisson02}, the phase-space and the distance-type
measures of nonclassicality were compared; it was shown in
\cite{Malbouisson02} that the distance-type measure is sensitive
to (non-Gaussian) superposition states (while $\tau _{m} $ is
maximum for such states) and, also, it results equivalent to the
phase-space measure introduced by Lee \cite{Lee91} for Gaussian
pure states. So, we can consider the distance-type measure as an
extension of that by Lee and in this report we will employ this
criterium for the measure of the nonclassicality of a field state.

At this point, a pertinent question emerges: Is it possible to
determine experimentally the nonclassical degree of a field state?
To answer this question, we will use a distance-type criterium to
characterize the nonclassicality of field states, and we shall
restrict our analysis to pure states. Metrics can be introduced in
the Hilbert space as functions of the fidelity, which corresponds
to the quantum-mechanical transition probability between two pure
states, ${\cal{F}}(\left| \Phi \right\rangle,\left| \Psi
\right\rangle) = \left| \left\langle \Phi | \Psi \right\rangle
\right|^{2}$. As examples, we mention the Bures-Uhlmann and the
Hilbert-Schmidt distances between two pure states ($\left| \Phi
\right\rangle $ and $\left| \Psi \right\rangle $), which are given
by
\begin{eqnarray}
d^{BU}(\left| \Phi \right\rangle,\left| \Psi \right\rangle ) & = &
\left( {2 - 2\,\sqrt{{\cal{F}}(\left| \Phi \right\rangle,\left|
\Psi
\right\rangle)} }\right)^{1/2} \, , \label{BU}\\
d^{HS}(\left| \Phi \right\rangle,\left| \Psi \right\rangle ) & = &
\sqrt {2 - 2\,{\cal{F}}(\left| \Phi \right\rangle,\left| \Psi
\right\rangle) } \, . \label{HS}
\end{eqnarray}
A distance-type measure of nonclassicality, for pure states, can
be defined as the minimum of any monotonically increasing function
of the distance between the state and an arbitrary coherent state
of the field mode. On these grounds, following
\cite{Malbouisson02}, we define the nonclassical degree of the
pure state $\left| \Psi \right\rangle $ as the minimum value of
one half of the squared Hilbert-Schmidt distance between $\left|
\Psi \right\rangle $ and an arbitrary coherent state  $\left|
\beta \right\rangle $, that is,
\begin{equation}
D_{\left| \Psi \right\rangle}=\min_{\{|\beta\rangle\}}\left[1 -
\left|\left\langle \beta | \Psi \right\rangle \right| ^{2}\right]
=1-\pi \max_{\left\{ \beta \in C\right\}}Q_{\left| \Psi
\right\rangle }\left( \beta \right) \, ,\label{profundidade}
\end{equation}
where $Q_{\left| \Psi \right\rangle }\left( \beta \right) $ is the
Husimi Q-function corresponding to the state $\left| \Psi
\right\rangle $. In other words, the quantity $D_{\left| \Psi
\right\rangle}$ defined in (\ref{profundidade}) will be used to
determine the degree of nonclassicality of the pure state $\left|
\Psi \right\rangle $, coherent states being taken as the most
classical between the quantum states of the field mode.

This measure of nonclassicality is slightly distinct from those in
[15,16]; besides being simpler, it also makes ease the comparison
with the nonclassical depth $\tau _{m}$ used in \cite{Lee91}, as
shown in \cite {Malbouisson02}. Note that $D_{\left| \alpha
\right\rangle}=0$ for coherent states, since $\max_{\{ \beta \in
{\bf C}\} } Q_{\left| \alpha \right\rangle }\left( \beta \right) =
\pi^{-1}$, as it should. On the other hand, for number states one
obtains $D_{\left| n \right\rangle}=1-n^{n}e^{-n}/n!$ showing
distinct nonclassical degree for different number states, the
upper bound ($D=1$) being reached in the limit $n\rightarrow
\infty $. This result differs from that emerging in the context of
phase-space measure \cite{Lee91} , where $\tau _{m}=1$ for all
number states $\left| n\right\rangle $, no matter the value of
$n$. At first glance, it may seem strange that a number state with
$n$ large is more nonclassical, in the sense of the distance
measure $D_{\left| n\right\rangle }$, than a state with smaller
$n$, say $\left| 1\right\rangle $. However, for a number state
$\left| n\right\rangle $, the expectation value of the
electromagnetic field vanishes identically while its energy is
proportional to $n$; clearly, such a state is more distant from
coherent states as larger $n$ is. Another distinction between
these two criteria comes from the fact that, while the
nonclassical depth $\tau _{m}$ introduced in \cite{Lee91} arises
from the {\it minimum} of the R-function, the nonclassical degree
$D_{\left| \Psi \right\rangle}$ of \cite{Malbouisson02}, for pure
states, arises from the {\it maximum} value of the Husimi
Q-function.

According to the Eq.(\ref{profundidade}), the experimental
determination of $D_{\left| \Psi \right\rangle}$ is obtained via
the Husimi Q-function. Then the question posed above can be
transposed to: ``How determining experimentally the Q-function?''
In a previous paper \cite{Baseia97}, we have proposed an
experimental arrangement to measure the Q-function. The strategy
follows the projection synthesis scheme proposed by Pegg-Barnett
\cite{Barnett96}. Accordingly, we can write $Q_{\left| \Psi
\right\rangle }\left( \beta \right) =\pi^{-1}\left\langle \beta
\right| \widehat{\rho }\left| \beta \right\rangle $, where
$\widehat{ \rho }=\left| \Psi \right\rangle \left\langle \Psi
\right| $ is the density operator describing the field whose
nonclassical degree is to be determined from
Eq.(\ref{profundidade}), and $\left| \beta \right\rangle $ stands
for a coherent state. It was shown in \cite{Baseia97} that
$Q_{\left| \Psi \right\rangle }\left( \beta \right) = \textrm{Tr}
(\widehat{\rho }\widehat{\Pi })$, where $\widehat{\Pi }=K$ $\left|
\beta \right\rangle \left\langle \beta \right| $, $K$ standing for
a constant. When we choose $\widehat{\Pi } =K$ $\left| \theta
\right\rangle \left\langle \theta \right| $, with $\left| \theta
\right\rangle $ being the phase-state, the method in
\cite{Barnett96} allows one to determine the phase-distribution
P$\left( \theta \right) $. Both cases require specific states used
as auxiliary reference fields, the {\it reciprocal binomial state}
in \cite{Barnett96} and the {\it complementary-coherent state} in
\cite {Baseia97}. Proposals for generation these two states in
travelling modes were recently suggested \cite{GRBS,sub1}. So, by
combining the experimental result for the Q-function with its
connection with the nonclassical degree $D_{\left| \Psi
\right\rangle}$ given in Eq.(\ref{profundidade}), one obtains the
experimental value of the quantity $D_{\left| \Psi
\right\rangle}$.

It is worth emphasizing that the above method concerns with states
of travelling modes of the quantized light field. What about
experimental method concerning with states of trapped fields
inside a high-Q cavity? In this case, there is an experimental
arrangement proposed by Lutterbach and Davidovich
\cite{Davidovich97} to determine the Wigner W-function describing
a stationary field inside a cavity. However, Q- and W-functions,
which are Gaussian convolutions of the Glauber-Sudarshan
P-function, are related by
\begin{equation}
Q\left( \beta \right) =\frac{2}{\pi }\int d^{2}\alpha W \left(
\alpha \right) e^{-2\left| \alpha -\beta \right| ^{2}},
\end{equation}
or, conversely, $W(\alpha )=\exp (-\frac{1}{2}\frac{\partial
}{\partial \alpha }\frac{\partial }{\partial \alpha ^{\ast
}})Q(\alpha )$. This equation, therefore, allows one to get the
Q-function via a Gaussian convolution of the Wigner function.
Alternatively, one can directly reconstruct the Q-function of an
initial state in a lossy cavity, as proposed in \cite{Hector}. We
should mention that other experimental proposals, also furnishing
the Wigner function, can be found in the literature for stationary
\cite{Bardroff96} and travelling \cite{Banaszek96} fields.

So far we have restricted our discussion of the degree of
nonclassicality to a single system, a mode of the electromagnetic
field. A natural question is then what is the nonclassical degree
of states of composite systems, for example two independent modes
($a$ and $b$) of the field. This will permit us to search for a
possible correspondence between the degrees of nonclassicality and
entanglement of field states.

To extend the distance-type measure of nonclassicality to
bipartite systems one has to choose a set of states as reference,
such states being considered as the most classical ones. A
possibility is to take the set of product states
$\left\{\left|\alpha ,\beta \right\rangle =
\left|\alpha\right\rangle_a \otimes
\left|\beta\right\rangle_b\right\}$, $\left|\alpha\right\rangle_a$
and $\left|\beta\right\rangle_b$ being coherent states of modes
$a$ and $b$ respectively, as the set of the most classical among
pure states of the bipartite system. In this way the definition
(\ref{profundidade}) can be generalized to
\begin{eqnarray}
D_{\left| \Psi \right\rangle_{ab}}&=&\min_{\{\left|\alpha ,
\beta\right\rangle\}}\left[1 - \left|\left\langle \alpha ,\beta |
\Psi \right\rangle_{ab} \right|
^{2}\right] \nonumber \\
 &=&1-\pi^2 \max_{\left\{ \alpha ,\beta \in
C\right\}}Q_{\left| \Psi \right\rangle_{ab} }\left( \alpha ,\beta
\right) \; ,\label{profundidade2}
\end{eqnarray}
where $Q_{\left| \Psi \right\rangle_{ab} }\left( \alpha ,\beta
\right)=\pi^{-2}\left\langle \alpha ,\beta |\hat{\rho}_{ab}
|\alpha ,\beta \right\rangle$ stands for the Husimi function of
the pure state $\left| \Psi \right\rangle_{ab}$. As naturally
expected, $D_{\left|\alpha\right\rangle_a \otimes
\left|\beta\right\rangle_b}=0$. Notice that if $\left| \Psi
\right\rangle_{ab}$ is a product state, that is $\left| \Psi
\right\rangle_{ab}=\left|\phi_1\right\rangle_a \otimes
\left|\phi_2 \right\rangle_b$, the Q-function is equal to the
product of the Husimi functions corresponding to the states of the
parts separately, $Q_{\left|\phi_1, \phi_2\right\rangle}(\alpha ,
\beta) = Q_{\left|\phi_1\right\rangle}(\alpha) \;
Q_{\left|\phi_2\right\rangle}(\beta)$. Thus, for product states,
the distance-type degree of nonclassicality can be expressed in
terms of the nonclassical degrees of the factor states, that is
\begin{equation}
D_{\left|\phi_{1}\right\rangle_a \otimes
\left|\phi_{2}\right\rangle_b}=D_{\left|\phi_{1}\right\rangle_a} +
D_{\left|\phi_{2}\right\rangle_b} -
D_{\left|\phi_{1}\right\rangle_a}
D_{\left|\phi_{2}\right\rangle_b} \, . \label{NDPS}
\end{equation}
In particular, one finds $D_{\left| 0\right\rangle_a \otimes
\left| n\right\rangle_b}=D_{\left| n\right\rangle_a \otimes \left|
0\right\rangle_b}=D_{\left| n \right\rangle}=1-n^{n}e^{-n}/n!$ and
$D_{\left| n\right\rangle_a \otimes \left| n\right\rangle_b} =
1-n^{2n}e^{-2n}/(n!)^2$.

Let us now consider the nonclassical degree for entangled states
of a bipartite system. Entanglement \cite{EPR} is widely believed
to be the fundamental trace distinguishing quantum mechanics from
classical mechanics, and it is crucial for aspects of quantum
information \cite{livro}, such as quantum teleportation, quantum
cryptography and quantum computation. One should then expect to
exist, somehow, a correspondence between entanglement and
nonclassicality of states.

To address this question consider, for simplicity, the families of
normalized states
\begin{equation}
\left| \Psi \right\rangle^{(\pm )}_{ab} =
\sqrt{\xi}\left|0,1\right\rangle \pm \sqrt{1 -
\xi}\left|1,0\right\rangle \label{ES}
\end{equation}
where $0\leq\xi\leq1$; these entangled states interpolate between
the one-photon product states $\left|1,0\right\rangle$ and
$\left|0,1\right\rangle$ of a bipartite system, which are the
limits $\xi = 0$ and $\xi = 1$ respectively. The degree of
entanglement, defined as the von Neumann entropy $\left[ \;
S_{N}(\hat{\rho})=-\textrm{Tr}(\hat{\rho} \ln{\hat{\rho}}) \;
\right]$ of either the reduced density matrix
$\hat{\rho}_{a}=\textrm{Tr}_{b}(\hat{\rho}_{ab})$ or
$\hat{\rho}_{b}$, for such states depend on $\xi$ and is given by
\begin{equation}
E(\xi)= -\left[ \xi\ln{\xi} + (1 - \xi)\ln{(1 - \xi )} \right] \,
, \label{entangle}
\end{equation}
irrespective of the sign taken in the superposition (\ref{ES}).
For $\xi=1$ or $\xi=0$, corresponding to the non-entangled states
$\left|0,1\right\rangle$ and $\left|1,0\right\rangle$
respectively, $E$ naturally vanishes while the maximum value of
$E(\xi)$ (namely $\ln 2$) is reached for $\xi = 1/2$, in which
case these entangled states look like the singlet or to one of the
triplet elements of the Bell's basis of the subspace of the
bipartite system (${\cal{H}}_{a} \otimes {\cal{H}}_{b}$) spanned
by $\left\{\left|0\right\rangle_{a} \left|0\right\rangle_{b} ,
\left|0\right\rangle_{a} \left|1\right\rangle_{b}
,\left|1\right\rangle_{a} \left|0\right\rangle_{b} ,
\left|1\right\rangle_{a} \left|1\right\rangle_{b} \right\}$. For
the states (\ref{ES}), the Husimi Q-function is given by
\begin{eqnarray}
Q^{(\pm )}_{\left| \Psi \right\rangle_{ab}}(\alpha ,\beta ;\xi) &
= & \frac{1}{\pi}\exp{\left[-(|\alpha|^2 +
|\beta|^2)\right]}\nonumber \\
 & & \times \left| \sqrt{\xi}\beta \pm \sqrt{1-\xi}\alpha \right|^2 \, ,
\label{Q}
\end{eqnarray}
and the nonclassical degree calculated by formula
(\ref{profundidade2}) is given by
\begin{equation}
D^{(\pm )}_{\left| \Psi \right\rangle_{ab}}(\xi) = 1 - e^{-1} \,
,\,\,\,\,\,\,\,\,\,\,\,\,\,\,\, 0\leq \xi \leq 1 \, ;
\label{const}
\end{equation}
that is, all members of both families of states (\ref{ES}) have
degree of nonclassicality equal to the nonclassical degree of the
states $\left|0,1\right\rangle$ and $\left|1,0\right\rangle$,
irrespective of the weights in the superpositions. Therefore, the
nonclassical degree of the states (\ref{ES}) is insensitive to
their degrees of entanglement.

We now consider the families of states
\begin{equation}
\left| \Phi \right\rangle^{(\pm )}_{ab} =
\sqrt{\xi}\left|0,0\right\rangle \pm \sqrt{1 -
\xi}\left|1,1\right\rangle \, , \label{ES2}
\end{equation}
with $0 \leq \xi \leq 1$, which interpolate between the
zero-photon state and a two-photon state (one in each mode) of the
bipartite system. The degree of entanglement of states (\ref{ES2})
is also given by Eq. (\ref{entangle}) and, again, the maximum
value of $E$ occurs when $\xi = 1/2$ for which states (\ref{ES2})
become the other Bell's states. For the states (\ref{ES2}), one
finds that the measure of nonclassicality (\ref{profundidade2})
leads to the same results for both $(+)$- and
$(-)$-superpositions, namely,
\begin{equation}
D^{(\pm )}_{\left| \Phi \right\rangle_{ab}}(\xi)=\left\{
\begin{array}{l}
1-(1-\xi )\exp \left[ -2\left( 1-\sqrt{\frac{\xi }{1-\xi }}\right)
\right],\;\xi \leq \frac{1}{2} \\
1-\xi
\;\;\;\;\;\;\;\;\;\;\;\;\;\;\;\;\;\;\;\;\;\;\;\;\;\;\;\;\;\;\;\;\;\;\;\;\;
\;\;\;\;\;\;\;\;,\;\xi \geq \frac{1}{2}
\end{array}
\right. \label{dm2}
\end{equation}
for $\xi$ within the interval $[0,1]$. One sees that $D^{(\pm
)}_{\left| \Phi \right\rangle_{ab}}(\xi)$ decreases from
$1-e^{-2}$ for $\xi = 0$ (which corresponds to the state
$\left|1,1\right\rangle$) to $0$ (the nonclassical degree of
$\left|0,0\right\rangle$) when $\xi = 1$, thus interpolating
monotonically between the nonclassical degrees of the constituting
states of the superpositions (\ref{ES2}); again, no correlation
between the degrees of nonclassicality and entanglement is found.
The degrees of nonclassicality and entanglement, for the families
of states (\ref{ES}) and (\ref{ES2}), are plotted in Fig.~1 as a
function of $\xi$.

\begin{figure}[h]
\includegraphics[{height=5.0cm,width=7.0cm}]{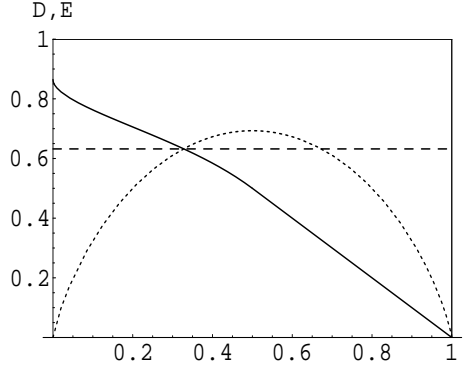}
\caption{Degrees of nonclassicality and entanglement, plotted as a
function of the parameter $\xi$, for the families of states
(\ref{ES}) and (\ref{ES2}): full and dashed lines refer to the
nonclassical degree of $\left| \Phi \right\rangle^{\pm}$ and
$\left| \Psi \right\rangle^{\pm}$ respectively; the dotted line
stands for the (common) degree of entanglement of these states.}
\end{figure}

The preceding analysis shows that entanglement is a quantum
property of states of bipartite systems which, as occur for others
nonclassical properties, does not alone determines the degree of
nonclassicality of a given state. However, for the families of
states considered, the distance-measure of nonclassicality
introduced is correlated with the nature of the photon statistics,
as indicated by the Mandel factor defined by
\begin{equation}
q_{\left| \psi \right\rangle} = \frac{\langle \psi | \hat{n}^2 |
\psi \rangle}{\langle \psi | \hat{n} | \psi \rangle} - \langle
\psi | \hat{n} | \psi \rangle - 1 \, . \label{MF}
\end{equation}
In fact, in ${\cal{H}}_{a} \otimes {\cal{H}}_{b}$, one defines
$\hat{n} = \hat{n}_a \otimes {\bf 1}_b + {\bf 1}_a \otimes
\hat{n}_b$, and using the closure relation ${\bf 1}_{ab} = \sum
|i\rangle_a |j\rangle_b\langle j |_b\langle i |_a$, one can easily
calculate $q_{\left| \psi \right\rangle}$ for products of number
states, $|n\rangle_a \otimes |m\rangle_b$, finding always $q=-1$;
such states are among the most sub-Poissonian states of a
bipartite system, like the number state $|n\rangle$ for a single
mode. The $q$-factor for superpositions of number-product states
can also naturally be evaluated and one finds, for the families
(\ref{ES}) and (\ref{ES2}) respectively,
\begin{eqnarray}
q^{\pm}_{\left| \Psi \right\rangle_{ab}}(\xi) & = & -1 \, ,
\label{qpsi} \\
q^{\pm}_{\left| \Phi \right\rangle_{ab}}(\xi) & = & 2\xi - 1 \, ;
\label{qphi}
\end{eqnarray}
one sees that $q^{\pm}_{\left| \Psi \right\rangle_{ab}}(\xi)$ take
the minimum value allowed, independently of $\xi$, while
$q^{\pm}_{\left| \Phi \right\rangle_{ab}}(\xi)$ vary from $-1$,
for $\xi = 0$, to $1$, when $\xi = 1$. Notice, in addition, that
the Bell states $\left| \Phi \right\rangle^{(\pm)}_{ab}(\xi =
1/2)$ are Poissonian. One concludes that the $q$-factor correlates
well with the nonclassical degree for the families of states
(\ref{ES}) and (\ref{ES2}).

In resume, we have discussed experimental routes for measuring the
degree of nonclassicality of field states describing a single
system. The extension of the notion of nonclassical degree (using
the distance-type criterium) for states of bipartite systems
allowed us to investigate possible connections between the degree
of nonclassicality and the degree of entanglement. As illustrated
in Fig. 1, such alluded correspondence does not exist for the
families of states $\left| \Psi \right\rangle^{\pm}$ and $\left|
\Phi \right\rangle^{\pm}$ considered here; $D^{(\pm )}_{\left|
\Psi \right\rangle_{ab}}$ does not change by varying $\xi$ and
$D^{(\pm )}_{\left| \Phi \right\rangle_{ab}}$ is a decreasing
function of $\xi$ in the whole interval $0\leq\xi\leq1$, while
$E(\xi)$ increases for $0\leq\xi\leq 1/2$ and decreases for $1/2
\leq\xi\leq 1$. On the other hand, a correlation was found between
the nonclassical degree and the Mandel factor for these families
of states. Many other states of a bipartite system can be analyzed
along these lines. Finally, we remark that the notion of
nonclassical degree we have introduced for bipartite states can be
easily extended to states of multi-partite systems, allowing a
comparison with the more general measure of entanglement recently
presented in Ref. \cite{Yukalov}. Such study is left for future
work.\\

This paper was partially supported by CNPq, FUNAPE (UFG) and
PRONEX, Brazilian agencies.

\end{document}